\documentclass{aastex631}
\usepackage{float, amsmath}

\newcommand{\eb}{\begin{equation}}
\newcommand{\ee}{\end{equation}}
\newcommand{\masyr}{mas yr$^{-1}$}
\newcommand{\uasyr}{$\mu$as yr$^{-1}$}

\definecolor{rkka}{RGB}{219,66,32}

\shorttitle{Differential spin of Hipparcos and Gaia proper motions}
\shortauthors{Makarov}

\begin{document}

\title{Differential proper motion spin of the Hipparcos and Gaia celestial reference frames}

\correspondingauthor{Valeri V. Makarov}
\email{valeri.makarov@gmail.com\\ valeri.v.makarov.civ@us.navy.mil}

\author[0000-0003-2336-7887]{Valeri V. Makarov}
\affiliation{U.S. Naval Observatory, 3450 Massachusetts Ave NW, Washington, DC 20392-5420, USA}

\begin{abstract}
The Hipparcos catalog provides the first epoch of the celestial reference frame (CRF) in the optical domain and serves as an indispensable tool to verify and improve the Gaia CRF for the brighter stars ($V<11$ mag) and to identify the elusive astrometric binary stars with dim or invisible companions, including long-period exoplanets. The systems of positions in Hipparcos and Gaia cannot be directly compared, because they refer to two different mean epochs. It is shown that the proper motion systems for carefully filtered common stars are not statistically consistent within the given formal errors. The vector field of proper motion differences is fitted with 126 vector spherical harmonics up to degree 7 revealing a global pattern at high signal-to-noise ratios, including the three terms of rigid rotation. The origin of the differential spin and other large harmonic terms is investigated by producing a similar decomposition of the Gaia$-$HG proper motion field, where HG stands for the long-term proper motions derived from the Hipparcos and Gaia DR3 positions, for the same sample of stars. Hipparcos proper motions emerge as the largest source of sky-correlated distortions of the multi-epoch optical CRF with a median value of $\sim 190$ \uasyr and a global spin of $\sim 226$ \uasyr, while the Hipparcos positions and Gaia EDR3 proper motions are explicitly consistent by construction at a level of $\sim 10$ \uasyr. The latter, however, include multiple distortions of higher degree, which should be taken into account in astrometric applications using the HG field.
\end{abstract}

\section{Introduction}
Publication of the Hipparcos catalog in 1997 marked a turning point in the development of astrometry as an astronomical discipline. The measurements of positions, trigonometric parallaxes, and proper motions had been taken to an unprecedented level of accuracy for about $120,000$ stars with magnitudes mostly brighter than $V=11$ mag. The new capabilities opened up a broad range of research opportunities \citep{2009aaat.book.....P}. In fundamental astrometry, a crucial shift of the definition of fundamental celestial reference frames away from a rotating, dynamically defined system with a precessing vernal equinox to a more abstract and quasi-inertial International Celestial Reference System was triggered \citep{2002A&A...392..341S}.
This shift became possible not only because of the higher accuracy of astrometric measurements but also due to the greater number of reference objects and a link to the radio reference system based on the VLBI measurements. Working in a stable, non-rotating reference frame provides new capabilities, including a more accurate description of the local Galactic kinematics, dynamics of the Solar system, and even some meaningful constraints and tests on the principles of cosmology and general relativity \citep{2021FrASS...8....9K}. The practical realization of such a system has a set of complications, however. As stars are not stationary on the sky, the reference frame performance is critically limited by the accuracy of proper motions. The highest accuracy is achieved close to the mean epoch of observations, but it quickly degrades away from that epoch because of the accumulating random and systematic errors of the proper motion system. Furthermore, a sizeable fraction of reference stars has nonlinear components of their apparent motion, which are not captured by the standard five-parameter astrometric model. The problem of differentiating ``good" and ``bad" astrometric stars is closely tied to the problem of correcting the systematic differences between the available proper motion system, which are both essential elements of the optical CRF maintenance.

Three kinds of proper motions are investigated in this paper. The Hipparcos proper motions (hereafter denoted HPM for brevity) are extracted directly from the Hipparcos astrometric catalog \citep{1997A&A...323L..49P}. These proper motions refer to the fixed reference epoch 1991.25. The Gaia DR3 proper motions (GPM) are also directly taken from the Gaia DR3 catalog \citep{2016A&A...595A...1G, 2021A&A...649A...1G}. These values refer to the fixed epoch 2016. The third kind of proper motions is calculated from the observed positions of common stars in Hipparcos and Gaia (denoted HGPM hereafter). They are long-term proper motions bridging the two reference epochs, which are almost statistically independent from the former two, and are of comparable precision to GPM.
The HGPM derived proper motions are needed to better identify astrometric binary stars that are likely to have physically perturbed astrometric data and to disentangle the possible causes of the detected systematic differences.

\section{Vetting the initial sample}
We begin with the original Hipparcos catalog, which includes $118,218$ entries. The first step is to select the objects with numerical values for proper motions, whose count is $117,955$. To cross-match this sample with Gaia DR3, the mean positions are extrapolated forward in time by 24.75 years onto the epoch 2016 using standard procedures of vectorial astrometry. The 2 by 2 covariance matrix of position is also transferred onto 2016 for reference purposes. The formal uncertainties of extrapolated Hipparcos positions are much greater than the mean-epoch uncertainties, because they include the HPM uncertainties multiplied by 24.75. We find that 20 stars, for example, have formal errors of declination in excess of $0.5\arcsec$ at 2016. This dictates a fairly accommodating radius of cross-match of $2\arcsec$ chosen for this study. Few statistical outliers or star with data perturbed by physical or perspective acceleration may be lost in this procedure, but the main priorities are accuracy and reliability, not completeness. 

This simple procedure yields $117,524$ Gaia counterparts. A few hundred missing stars are predominantly known double or multiple stars with strongly perturbed or even incorrect Hipparcos astrometry \citep{2000A&AS..144...45F}, with a smaller addition of very bright stars missing in Gaia. An interesting example of the former is HIP 1242, a nearby M4V red dwarf, whose extrapolated position is $22.126\arcsec$ away from the observed Gaia position, indicating a perturbation of 80.8 \masyr\ in HPM. Known binary stars are useless for this study, and the next step is to filter out all stars marked with multiplicity flags X (stochastic solutions), O (orbital solutions), and C (component solutions). For a short description of binarity categories and their impact on astrometry of bright stars, cf. \citep{2022AJ....164...36Z}. The number of objects in the sample drops to $102,828$. After removing all Gaia stars without proper motion data, the sample shrinks to $102,549$.

\section{Computing long-term HGPM}
\label{hgpm.sec}

Using the well-known formulae of spherical astronomy \citep{1985spas.book.....G}, the 3D position vectors $\boldsymbol r_{\rm H}$ and $\boldsymbol r_{\rm G}$ are computed from the mean equatorial coordinates given in the Hipparcos and Gaia EDR3 catalogs, respectively. These unit vectors define the mean positions of each star on the celestial sphere. The 3D proper motion vector of the HGPM type can be approximately computed as
\eb
\boldsymbol \mu_{\rm HG}= (\boldsymbol{r}_{\rm G}-\boldsymbol{r}_{\rm H})/24.75,
\ee
in radians per year, where 24.75 is the epoch difference in years. We note that this is a geometric approximation of the more exact formula, which substitutes the arc length between the two positions with the chord length. The introduced bias is negligible even for the fastest stars, on a short time scale. This long-term vector should be referred to the position where the observed proper motion to be compared with is defined. In our case, this position is $\boldsymbol{r}_{\rm G}$. 

The formal covariance of the HG proper motion vector is readily computed from the corresponding position matrices $\boldsymbol{C}_{r\rm H}$ and $\boldsymbol{C}_{r\rm G}$:
\eb 
\boldsymbol{C}_{\mu\rm HG}=(\boldsymbol{C}_{r\rm H}+\boldsymbol{C}_{r\rm G})/24.75^2.
\ee 
This simple formula is correct as long as the observed positions in the two catalogs are statistically independent.

\section{GPM minus HGPM differences and astrometric binary stars}

Once a HGPM vector and its formal covariance are known for each star, it is straightforward to compute the proper motion differences 
\eb 
\boldsymbol \delta_{\rm G-HG}=\boldsymbol \mu_{\rm G}-\boldsymbol \mu_{\rm HG}.
\ee 
The corresponding formal uncertainty is represented by the covariance matrix
\eb 
\boldsymbol{C}_{\mu\rm G-HG}=\boldsymbol{C}_{\mu\rm HG}+\boldsymbol{C}_{\mu\rm G}
\label{cov.eq}
\ee 
which is a simplifying approximation, because it does not include the correlation coefficients of the Gaia mean positions and proper motions. It is acceptable to ignore these off-diagonal elements because their contribution is small compared to the other terms.

\begin{figure*}
    \includegraphics[width=0.48 \textwidth]{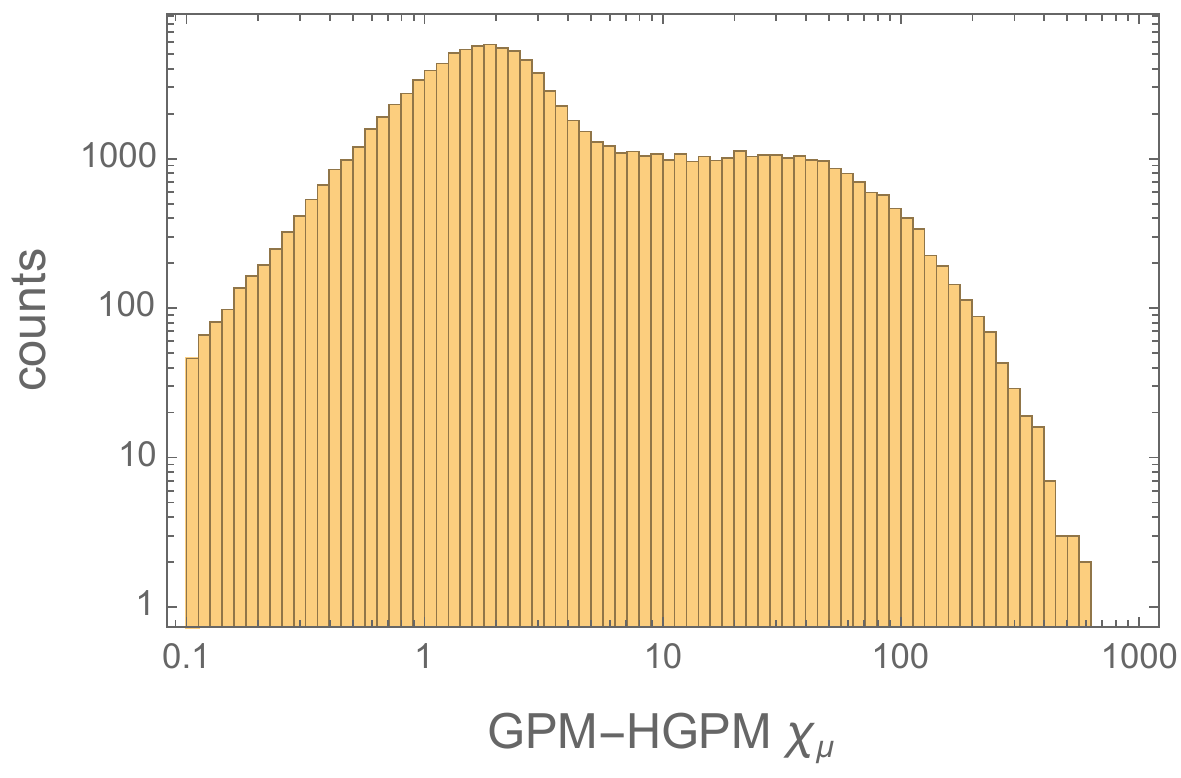}
    \caption{Histogram of $\chi$-values of GPM$-$HGPM differences for $102,549$ stars
    in Hipparcos and Gaia DR3. Note the logarithmic scale on both axes, which, coupled with bins of equal width, emphasize the surplus of outliers and shifts the peak away from the expected modal value of 1.}
    \label{chimu.fig}
\end{figure*}

Using these derived data, a $\chi$-value can be computed for each proper motion difference as
\eb 
\chi_\mu=\left(\boldsymbol \delta_{\rm G-HG}^T\,\boldsymbol{C}_{\mu\rm G-HG}^{-1}\,
\boldsymbol \delta_{\rm G-HG}\right)^\frac{1}{2}
\ee 
If the random errors are normally distributed with a zero mean and the formal covariances truly represent their uncertainty, the bivariate $\chi$ statistic should be
$\chi$-distributed with 2 degrees of freedom (which is the same as Rayleigh-distributed with a scale parameter of 1). Fig. \ref{chimu.fig} shows the actual histogram of $\chi$-values in log--log axes. We find a huge surplus of strongly perturbed vectors whose length is much greater than the range of expected values. To illustrate the scale of this feature, consider that the
CDF$[\chi(2)]$ at 5 is 0.999996, so that less than one star in the sample is expected to exceed this value, while we find many thousands. The mean of $\chi(2)$ is approximately 0.707. If we assign the hump at $\chi>5$ to previously unknown astrometric binaries with apparent acceleration of motion \citep{1999A&A...346..675W, 2003AN....324..419K, 2005AJ....129.2420M}, the core distribution may be perturbed by smaller effects. For the nearest stars, the orbital perturbation from Jupiter-mass long-period exoplanets can be large enough to be confidently detected in a similar differential analysis of long-term and Gaia proper motions \citep{2021RNAAS...5..155M}. However, there is no strong correlation of the higher $\chi$-values with parallax except for the smallest parallaxes (most distant stars). In this paper, we investigate the possibility that this error overhead comes from large-scale correlated errors in GPM or HGPM.

Using the results shown in Fig. \ref{chimu.fig}, we further clean the working sample of Hipparcos--Gaia stars by removing the obviously perturbed (presumably, by astrometric binarity) stars with $\chi>5$. This threshold value is close to the inflection point in the histogram, which is deemed to balance the risks of removing too many valid data points and accepting too many binaries. This cut reduces the number of objects to $75,686$, i.e., a reduction by 26\%. 

\section{Fitting the large-scale differential proper motion field}

Having selected a set of astrometrically reliable stars in the core distribution, we can investigate the global pattern of sky-correlated fields, which are called ``distortions" hereafter. An advanced vector spherical harmonic (VSH) decomposition technique is employed in this study. Originally proposed for astrometric analysis of large catalogs by \citet{2004ASPC..316..230V}, this techniques is best suited to reveal distortions on the large to medium spatial scales, because the VSHs are intrinsically orthogonal in the Hilbert space of vector-valued functions, which can minimize the unwanted correlations of the fitting parameters. It has been applied to the system of Hipparcos proper motions as a whole to estimate the local parameters of Galactic kinematics \citep{2007AJ....134..367M, 2015AJ....149..129M}, as well as to compare two astrometric catalogs \citep{2011AstL...37..874V, 2012A&A...547A..59M} and to improve the tie between the radio and optical celestial reference frames \citep{2019gaia.confE..25M,2021AJ....161..289M}.  Global astrometric missions can be modeled and optimized using VSH fields as free parameters instead of the traditional astrometric unknowns for each object \citep{2012AJ....144...22M}.

The complete infinite expansion of a real-valued field $\boldsymbol u$ is
\eb
\begin{split}
\boldsymbol u (\alpha,\delta)= \sum_{l=1}^{+\infty} [ & c_{0l0}{\bf EVSH}_{0l0}(\alpha,\delta)+ d_{0l0}{\bf MVSH}_{0l0}(\alpha,\delta) \\
&+\sum_{k=1}^2 \sum_{m=1}^l\; c_{klm}{\bf EVSH}_{klm}(\alpha,\delta) +
d_{klm}{\bf MVSH}_{klm}(\alpha,\delta) ]
\end{split}
\label{vsh.eq}
\ee
where
\begin{eqnarray}
{\bf EVSH}_{0l0}&=&\boldsymbol S_l^0 \nonumber \\
{\bf MVSH}_{0l0}&=&\boldsymbol T_l^0 \nonumber \\
{\bf EVSH}_{1lm}&=&{\mathfrak{Re}}\left[\boldsymbol S_l^m\right] \nonumber \\
{\bf MVSH}_{1lm}&=&\mathfrak{Re} \left[\boldsymbol T_l^m\right] \nonumber \\
{\bf EVSH}_{2lm}&=&\mathfrak{Im}\left[\boldsymbol S_l^m\right] \nonumber \\
{\bf MVSH}_{2lm}&=&\mathfrak{Im}\left[\boldsymbol T_l^m\right],
\end{eqnarray}
with $\boldsymbol S_l^m$ and $\boldsymbol T_l^m$ being the poloidal (electric) and toroidal (magnetic) complex-valued VSHs. The $k$ index is to distinguish the two kinds of VSHs of nonzero order ($m>0$) coming from the real and imaginary parts of the complex-valued functions, which have orthogonal phases in the Fourier parts. $l$ is the degree, and $m$ is the order of a VSH. 

In practical applications, these VSH expansions are truncated at a certain degree $L$. The number of independent and mutually orthogonal harmonics for each degree $l$ is $2(1+2l)$, so that the total number of terms in a truncated expansion is $N_{\rm VSH}=2L(2+L)$. Note that the number of fitting terms quadratically increases with $L$. Using Wolfram Mathematica, analytical expressions can be quickly found for VSHs of practically any degree and order. For example,
\eb
{\bf MVSH}_{296}=\left\{\begin{aligned} &
\frac{3}{1024}\sqrt{\frac{40755}{\pi}}\sin(6 \alpha)\cos^5\delta (65 - 108 \cos(2 \delta) + 51 \cos(4 \delta) )\\ &
\frac{3}{256}\sqrt{\frac{40755}{\pi}}\cos(6 \alpha)\cos^5\delta (39 \sin \delta - 17 \sin(3 \delta) )
\end{aligned}\right\}
\ee
Note that each VSH is a 2-vector field composed of the projections onto $\boldsymbol\tau_\alpha$ and $\boldsymbol\tau_\delta$, as described in \citep{2007AJ....134..367M}.

The unknown coefficients $c_{klm}$ and $d_{klm}$ in expansion (\ref{vsh.eq}) can be determined in a weighted least-squares adjustment. Two weighting schemes have been proposed in the literature, an ``optimal" weighting based on the formal covariances of the observed vectors, and an iterative ``empirical" scheme based on the post-fit residuals for each object or data point. The latter provides more reliable and stable results when the input data are heavily perturbed and the formal errors do not capture the scale of this perturbation. For example, a general proper motion field includes stars with their peculiar physical motion that may deviate from the systemic Galactic motion by great amounts. For this study, the former scheme is adopted, because the sample has been filtered for astrometric binaries with the largest perturbations beyond the formal error budget. 

The problem of fitting a truncated VSH expansion for a given vector filed can be written as
\eb
\boldsymbol{A}\;\boldsymbol{x}=\boldsymbol{y},
\label{a.eq}
\ee
where $\boldsymbol{A}$ is the design matrix with $2L(2+L)$ columns, $\boldsymbol{x}$ is the vector of unknown coefficients $c_{klm}$ and $d_{klm}$, and $\boldsymbol{y}$ is the vector of observations. The order of VSH functions in the columns of $\boldsymbol{A}$, which is identical to the order of the unknown coefficients, can be arbitrary, in principle. It is important, however, to avoid confusion with the adopted ordering scheme in the interpretation of the results. Also, unlike a regular design matrix represented by a 2D array, $\boldsymbol{A}$ has dimensions $N_{\rm obs}\times2L(2+L)
\times 2$, i.e., a 3D array. The matrix multiplication involves the inner scalar product of 2-vectors rather than the regular multiplication of scalars. Consequently, the weight matrix can be represented as a $N_{\rm obs}\times2L(2+L)
\times 2\times2$ array, although this may not be practical for problems with a large number of data points $N_{\rm obs}$. The optimal weight for each data point is computed as the inverse of the $2\times2$ formal covariance matrix, e.g., per Eq. \ref{cov.eq}. We note that the standard procedure of pre-multiplying separately both the design matrix $\boldsymbol{A}$ and the right-hand part $\boldsymbol{y}$ with a weight matrix would also require computing the matrix square root of the inverse, and placing these blocks in a tridiagonal sparse matrix. It is more practical, however, to construct the normal equations right away folding the extra dimensions:
\eb
(\boldsymbol{A}^T \boldsymbol{W} \boldsymbol{A})\,\widehat{\boldsymbol{x}}\,=\,
(\boldsymbol{A}^T \boldsymbol{W} \,\boldsymbol{y}),
\label{norm.eq}
\ee 
and solving them in the usual way by computing the covariance matrix of the solution vector $\boldsymbol{x}$:
\eb 
\boldsymbol{C}_{\widehat{\boldsymbol{x}}}=(\boldsymbol{A}^T \boldsymbol{W} \boldsymbol{A})^{-1}.
\label{covx.eq}
\ee 

\section{VSH decomposition of GPM $-$ HGPM differences}

A weighted VSH solution with $L=7$ (126 fitting functions) was computed for the $75,686$ Hipparcos--Gaia stars without large astrometric perturbations (Section \ref{hgpm.sec}). To reduce the computing time, which is significant due to the multi-dimensional nature of the problem and non-standard way of weighting, I used a cell-averaging technique. The sky is divided into 4584 cells of approximately equal area of $3\times 3$ deg$^2$ in belts of equal declination \citep[cf. a similar technique used by ][ to compare the positional systems of the radio and optical reference frames]{2021MNRAS.506.5540M}. The mean proper motion vector (GPM $-$ HGPM) is computed from all stars located within each cell using the formal covariance matrices $\boldsymbol{c}_i$ for weights:
\eb 
\langle \boldsymbol{v} \rangle = \sum_i \boldsymbol{w}_i\, \boldsymbol{v}_i,
\ee 
where 
\eb 
\boldsymbol{w}_i = \left(\sum_i \boldsymbol{c}_i^{-1} \right)^{-1}\,\boldsymbol{c}_i^{-1}.
\ee 
The first multiplier in this formula is obviously the formal covariance of the weighted mean vector $\langle \boldsymbol{v} \rangle$, which is subsequently used for computing the weights in the VSH adjustment. The geometric centers of the cells yield the coordinates of reference points (origins of $\langle \boldsymbol{v} \rangle$). We note that the size of each cell is much smaller than the characteristic wavelength of the fitted VSH, which precludes any artificial smoothing of the resulting vector field.

The resulting system of condition equations is $4584\times 126\times 2$, which is quite manageable on a regular laptop. Solving it produces a vector $\widehat{\boldsymbol{x}}$ of 126 coefficients $c_{klm}$ and $d_{klm}$ ordered in the same way as the corresponding functions in the columns of the design matrix. The units of these coefficients are \masyr\ in our case. It is not possible to tell if the fitted VSH are statistically significant just from the computed values of the coefficients but the covariance matrix $\boldsymbol{C}_{\widehat{\boldsymbol{x}}}$ (Eq. \ref{covx.eq}) contains the formal variances in the diagonal, which can be used to compute a statistic variable analogous to the signal-to-noise ratio (S/N):
\eb 
f_j=\widehat{x}_j/\sqrt{C_{jj}}.
\label{sn.eq}
\ee 
The fitted differential proper motion field is simply computed as
\eb 
\widehat{\boldsymbol{u}}=\boldsymbol A\;\widehat{\boldsymbol{x}}.
\ee 

\begin{figure*}
    \includegraphics[width=0.99 \textwidth]{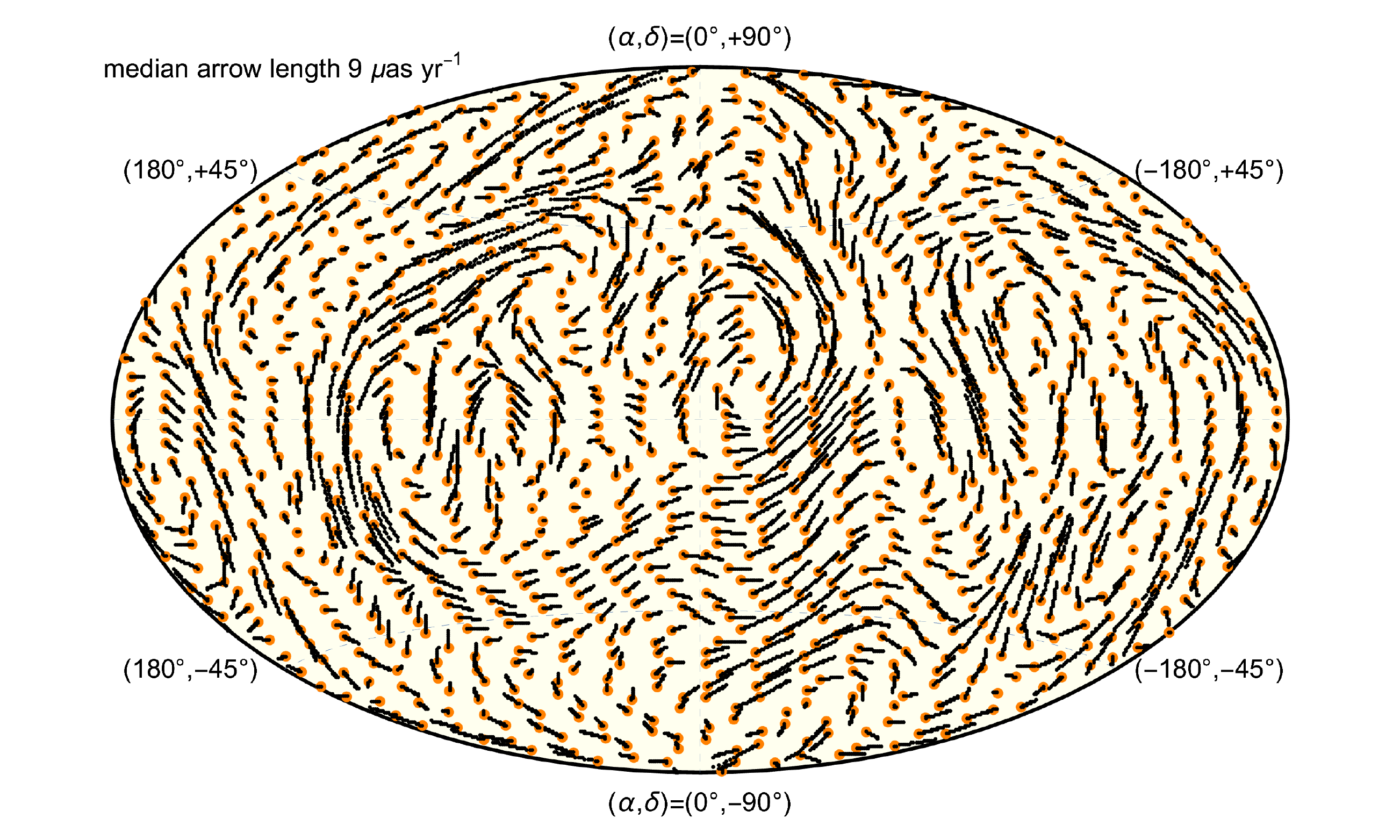}
    \caption{Aitoff projection in equatorial coordinates of the GPM $-$ HGPM differential vector field fitted with 126 VSH up to degree 7. The orange dots indicate the origins of vectors, not their heads.}
    \label{pro1.fig}
\end{figure*}

The results for the GPM $-$ HGPM differential proper motion field are shown in the graphical form in Fig. \ref{pro1.fig}. The arbitrarily scaled vectors $\widehat{\boldsymbol{u}}$ are shown in the Aitoff sky projection of the equatorial coordinate system. The orange dots indicate the positions of the origins of these vectors, i.e., the centers of the averaging cells. We find a complex pattern of stream-like motion on the medium spatial scale. A sequence of local vortices seems to be present in the broad equatorial zone, which may be aligned with the ecliptic hinting at an origin in the scanning law of the space missions.

\startlongtable
\begin{deluxetable*}{C C RCC c}
\tabletypesize{\footnotesize}
\tablecaption{Significant VSH terms in GPM $-$ HGPM expansion}
\label{t1.tab}
\tablehead{
\colhead{\rm number}  & \colhead{VSH id}   & \colhead{value \masyr} & \colhead{$\sigma$ \masyr} & \colhead{S/N}  \\ } 
\startdata
 1 & \{\text{mag},0,1,0\} & 0.004724 & 0.000453 & 10.427304 \\
 2 & \{\text{mag},1,1,1\} & 0.010706 & 0.000469 & 22.844049 \\
 3 & \{\text{mag},2,1,1\} & 0.008658 & 0.000504 & 17.190345 \\
 5 & \{\text{ele},1,1,1\} & 0.003797 & 0.000561 & 6.765568 \\
 7 & \{\text{mag},0,2,0\} & 0.002383 & 0.000218 & 10.912859 \\
 10 & \{\text{ele},0,2,0\} & -0.001156 & 0.000188 & -6.150764 \\
 13 & \{\text{mag},1,2,2\} & -0.001546 & 0.000295 & -5.235877 \\
 18 & \{\text{mag},1,3,1\} & -0.003644 & 0.000226 & -16.146184 \\
 19 & \{\text{mag},2,3,1\} & -0.001744 & 0.000217 & -8.027109 \\
 20 & \{\text{ele},0,3,0\} & -0.000856 & 0.000135 & -6.355415 \\
 23 & \{\text{mag},1,3,2\} & 0.002660 & 0.000212 & 12.573519 \\
 28 & \{\text{mag},2,3,3\} & -0.001718 & 0.000207 & -8.312381 \\
 33 & \{\text{mag},2,4,1\} & 0.001367 & 0.000170 & 8.054794 \\
 36 & \{\text{ele},2,4,1\} & -0.000839 & 0.000145 & -5.777135 \\
 45 & \{\text{mag},1,4,4\} & -0.001138 & 0.000165 & -6.880918 \\
 51 & \{\text{mag},2,5,1\} & -0.001364 & 0.000138 & -9.904411 \\
 56 & \{\text{mag},2,5,2\} & -0.000742 & 0.000135 & -5.483628 \\
 60 & \{\text{mag},2,5,3\} & 0.001629 & 0.000135 & 12.106997 \\
 63 & \{\text{mag},1,5,4\} & -0.001842 & 0.000131 & -14.014721 \\
 64 & \{\text{mag},2,5,4\} & -0.001196 & 0.000138 & -8.673065 \\
 85 & \{\text{mag},1,6,4\} & 0.000675 & 0.000109 & 6.183531 \\
 87 & \{\text{ele},1,6,4\} & 0.001078 & 0.000109 & 9.860958 \\
 91 & \{\text{ele},1,6,5\} & 0.001220 & 0.000117 & 10.390628 \\
 99 & \{\text{mag},2,7,1\} & 0.000572 & 0.000096 & 5.932522 \\
 104 & \{\text{mag},2,7,2\} & -0.001417 & 0.000096 & -14.713398 \\
 107 & \{\text{mag},1,7,3\} & 0.001130 & 0.000095 & 11.887134 \\
 109 & \{\text{ele},1,7,3\} & 0.000481 & 0.000087 & 5.537172 \\
 112 & \{\text{mag},2,7,4\} & 0.001343 & 0.000094 & 14.224257 \\
 115 & \{\text{mag},1,7,5\} & -0.000500 & 0.000094 & -5.336129 \\
 116 & \{\text{mag},2,7,5\} & -0.001529 & 0.000096 & -15.847110 \\
 120 & \{\text{mag},2,7,6\} & -0.000842 & 0.000098 & -8.608552 \\
 123 & \{\text{mag},1,7,7\} & -0.001083 & 0.000100 & -10.792573 \\
 124 & \{\text{mag},2,7,7\} & -0.002405 & 0.000097 & -24.765939 \\
\enddata
 \tablecomments{Columns description: 1) running number of VSH in the accepted ordering scheme; 2) VSH identification: {\tt mag} for magnetic, {\tt ele} for electric, followed by the real (1) or imaginary (2) index, except for $m=0$ terms, followed by degree $l$ and order $m$;
 3) fitted coefficient; 4) standard deviation of fitted value; 5) significance of fitted value.}
\end{deluxetable*}

Table \ref{t1.tab} lists all VSH terms with S/N values (Eq. \ref{sn.eq}) greater than 5, which are deemed statistically significant. The elevated S/N threshold accounts for the sample distribution of $\chi_{\boldsymbol \mu}$ values (Fig. \ref{chimu.fig}), which shows evidence of a general unmodeled perturbation of the input data apart from the presence of numerous astrometric binaries. With this conservative criterion, we find 33 (out of 126) statistically significant harmonics. At a higher confidence level, there are 15 VSH terms with S/N$>10$. This set is dominated by the harmonics of the magnetic type, which confirms the visual impression of a series of vortices in the graph. The VSH spectrum is fairly ``flat", in the sense that large terms continue to appear at high degrees---indeed, the most significant term is number 124, which is {\bf MVSH}$_{277} = \left\{\frac{21}{64} \sqrt{\frac{715}{2 \pi }} \sin (7 \alpha ) \sin (\delta ) \cos ^6(\delta ),\frac{21}{64} \sqrt{\frac{715}{2 \pi }} \cos (7 \alpha ) \cos ^6(\delta )\right\}$. The existence of such medium-scale distortions may be attributable to the systematics of the Gaia proper motion solution, because a similar pattern has been found with a different technique within the footprint of the Sloan Digital Sky Survey \citep{2022ApJ...927L...4M}. Among the low-degree terms, the first three magnetic harmonics make a prominent group with their relatively large amplitudes and high S/N estimates. These represent the rigid rotation of the entire differential vector field, i.e., the global spin of the GPM field with respect to the HGPM filed.

\section{VSH decomposition of the GPM $-$ HPM vector field}

Using the same sample of $75,686$ astrometrically stable star (with small or moderate accelerations) and the same procedure, a weighted least-squares fit is performed for the GPM $-$ HPM field, i.e., the direct difference between the Gaia and Hipparcos proper motions for common stars. The transformation of proper motion vectors between the two epochs is omitted for this study, because the incurred errors are much smaller than the observational error of the Hipparcos data. The formal covariance $\boldsymbol C_{\mu {\rm G-H}}$ is then simply the sum of the proper motion covariances given in the two catalogs. The two components are incompatible, because the formal uncertainty in Hipparcos is 25--50 times that in EDR3. The same cell-averaging technique and weighting scheme are used again.

\begin{figure*}
    \includegraphics[width=0.99 \textwidth]{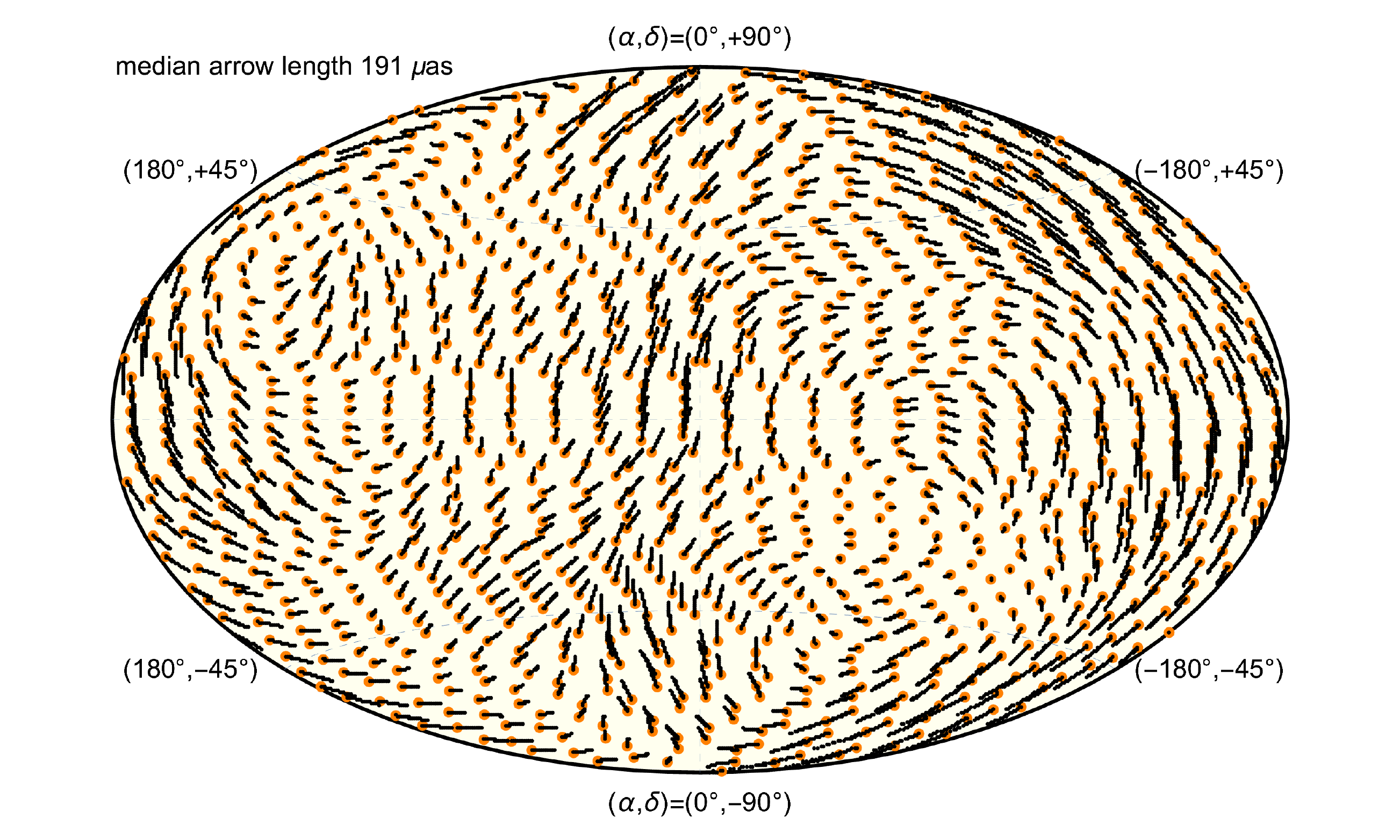}
    \caption{Aitoff projection in equatorial coordinates of the GPM $-$ HPM differential vector field fitted with 126 VSH up to degree 7. The orange dots indicate the origins of vectors, not their heads.}
    \label{pro2.fig}
\end{figure*}

Fig. \ref{pro2.fig} shows the sky projection of the GPM $-$ HPM differential proper motion field with 126 VSHs up to degree $L=7$. Comparing this pattern with Fig. \ref{pro1.fig} for the GPM $-$ HGPM field, we find much greater distortions in absolute values with a median length of 191 \uasyr, which is approximately 20 times larger. The patterns also looks much smoother, dominated by low-degree harmonics compounding to large--scale flows over substantial parts of the sphere. Only a couple of local vortices are apparent. Perhaps, the most conspicuous feature is a large meridional southward stream at $\alpha$ between $170\degr$ and $220\degr$. The largest fitted vector is found at \{RA, Decl.\}$=\{253\degr,47\degr\}$, which is $\{-256,-288\}$ \uasyr. 

\startlongtable
\begin{deluxetable*}{C C RCC c}
\tabletypesize{\footnotesize}
\tablecaption{Significant VSH terms in GPM $-$ HPM expansion}
\label{t2.tab}
\tablehead{
\colhead{\rm number}  & \colhead{VSH id}   & \colhead{value \masyr} & \colhead{$\sigma$ \masyr} & \colhead{S/N}  \\ } 
\startdata
 1 & \{\text{mag},0,1,0\} & 0.205912 & 0.010039 & 20.511861 \\
 2 & \{\text{mag},1,1,1\} & -0.291418 & 0.010249 & -28.434594 \\
 3 & \{\text{mag},2,1,1\} & 0.506941 & 0.010447 & 48.527226 \\
 24 & \{\text{mag},2,3,2\} & -0.028656 & 0.004942 & -5.798584 \\
 74 & \{\text{ele},0,6,0\} & 0.010609 & 0.001483 & 7.155928 \\
 75 & \{\text{ele},1,6,1\} & 0.017262 & 0.002048 & 8.428981 \\
 87 & \{\text{ele},1,6,4\} & -0.019891 & 0.002332 & -8.530531 \\
 105 & \{\text{ele},1,7,2\} & -0.009411 & 0.001768 & -5.323259 \\
 112 & \{\text{mag},2,7,4\} & -0.010601 & 0.002007 & -5.280996 \\
\enddata
 \tablecomments{Columns description: 1) running number of VSH in the accepted ordering scheme; 2) VSH identification: {\tt mag} for magnetic, {\tt ele} for electric, followed by the real (1) or imaginary (2) index, except for $m=0$ terms, followed by degree $l$ and order $m$;
 3) fitted coefficient; 4) standard deviation of fitted value; 5) significance of fitted value.}
\end{deluxetable*}

Table \ref{t2.tab} lists all VSH terms with S/N values $f_j$ greater than 5 in absolute value. In comparison with Table \ref{t1.tab}, we find much fewer very significant terms in the decomposition, 9 against 33. Obviously, the larger formal uncertainties of this result do not allow us to detect more signals with confidence. However, we note that the largest terms appear in the first triplet of first--degree magnetic harmonics, that is, the rigid spin of the entire proper motion field. This defines the overall appearance of a smooth flowing pattern in Fig. \ref{pro2.fig}. Using the definitions of the spin velocity projections $\omega_i$, $i=X,Y,Z$, in \citep{2021A&A...649A.124C}, the following relations are established:
\begin{eqnarray}
\omega_X &=& -\frac{1}{2}\sqrt{\frac{3}{2\pi}}d_{111} \\ \nonumber
\omega_Y &=& -\frac{1}{2}\sqrt{\frac{3}{2\pi}}d_{211} \\ \nonumber
\omega_Z &=& -\frac{1}{2}\sqrt{\frac{3}{\pi}}d_{010} 
\end{eqnarray}
The fitted values of the GPM $-$ HPM field are $\omega_X=+100.7\pm 3.5$ \uasyr, $\omega_Y=-175.1\pm 3.6$ \uasyr, and  $\omega_Z=-100.6\pm 4.9$ \uasyr.

\section{Discussion}
\citet{2021A&A...649A.124C}, using resolved binary stars and open clusters, compared the general spin of the Gaia EDR3 proper motion system for brighter and fainter stars and found significant magnitude-dependent differences for stars brighter than $G=13$ mag. These are believed to represent systematic errors in Gaia proper motions for the brighter stars, which were observed and processed in a different regime, while the faint stars are much better adjusted to the fixed quasars and AGNs. In the range of magnitudes relevant for Hipparcos, the spin parameters $\omega_{\rm bright}- \omega_{\rm faint}$ come up to $+60$ \uasyr\ in the $\omega_Y$ parameter and $+20$ \uasyr\ in $\omega_X$, with $\omega_Z$ being closer to zero. The present study has revealed numerically much smaller spins of the GPM $-$ HGPM differential proper motion field with estimated values $\omega_X=-2.3\pm 0.2$ \uasyr, $\omega_Y=-3.7\pm 0.2$ \uasyr, and $\omega_Z=-3.0\pm 0.2$ \uasyr. The apparent discrepancy of these results is explained by a feature in the production of the Gaia DR3 astrometric solution for stars with $G<13$ mag \citep[][section 4.5]{2021A&A...649A...2L}. A rigid spin was introduced in the solution for brighter stars with the purpose of adjusting this proper motion system to the positional systems of Hipparcos and Gaia, effectively setting the spin to that of the HGPM field. The introduced error was estimated to be within 24 \uasyr\ from the assumed positional alignment of $\sim 600$ $\mu$as of Hipparcos to the ICRS \citep{1997A&A...323..620K}. The authors emphasized that only the rigid spin of the GPM system had been fixed. The other higher-degree VSH terms found in this paper should therefore reflect the actual performance of the GPM and HGPM systems. Furthermore, the rigid spin components, small as they are, are still highly significant, which indicates that the adjustment of the GPM system was not quite efficient.

Effectively, the global rotations of the Hipparcos system of positions have been transferred with diluted magnitudes into the Gaia proper motion system, but not the higher-order distortions. This probably explains the complex pattern of the fitted GPM $-$ HGPM filed (Fig. \ref{pro1.fig}). Can these data be used for the search of faint and close companions with the $\Delta\mu$-method? Obviously, when the detected signal is much greater than the value of distortion ($\sim 10$ \uasyr), these errors can be ignored. But selecting candidate exoplanets may require a careful approach, including possible subtraction of the fitted VSH field. The much larger, magnitude-dependent spin components should be taken into account only if absolute proper motions referred to the ICRS are required, for example, in studies of Galactic rotation.

Our results for the GPM $-$ HPM differential field paint a very different picture (Fig. \ref{pro2.fig} and Table \ref{t2.tab}). The magnitudes of fitted VSH terms on exactly the same sample of stars are roughly 20 times greater, and the emerging pattern is that of smooth large-scale flows, dominated by the three magnetic harmonics of rigid spin. The formal errors of the VSH coefficients are also $\sim 20$ times larger driven by the formal uncertainties of the Hipparcos proper motions. These distortions are larger in magnitude than the estimated systematic errors of Gaia proper motions at $G<11$ mag. The largest term is the spin component $\omega_Y$, which amounts to $-175.1\pm 3.6$ \uasyr. Formally, this differential spin can be represented as angular acceleration of the entire optical reference frame (i.e., positional system of brighter stars) around the vector pointing at $\{\alpha,\delta\}=\{299.9\degr, -26.5\degr\}$ of $(225.7\pm 7.0)/24.75$ $\mu$as yr$^{-2}$. Over the 24.75 yr between the mean epochs of Hipparcos and Gaia DR3, this acceleration accounts to a positional misalignment of $\sim2.6$ mas. The six significant harmonics of higher degree combine into a patchy pattern with some areas of nearly zero distortion adjacent to large fields of much greater differences, up to 4 mas over the same time interval.

\section{Conclusions}
The formally most accurate information is obtained for the GPM $-$ HGPM differential field, which confirms that the explicit spin between these proper motion systems is numerically small by construction. The rigid spin expressed through the first three magnetic VSHs comes up to a few \uasyr. A nonzero result in this case is explained by a different sample and adjustment technique used in the Gaia pipeline. On the other hand, we know that the Gaia DR3 proper motion system of brighter stars suffers from much greater systematic errors. Furthermore, the modal value of proper motion differences is about 70 \uasyr. The fitted  high-degree VSH field cannot explain the observed distribution. Therefore, there should be an additional and much stronger source of perturbation in the GPM $-$ HGPM data. The possibilities are limited to small-scale correlated errors (which are not captured by the VSH fitting to $L=7$) in Gaia proper motions or in Hipparcos positions. The latter contribute more, even diluted over 24.75 yr, to the formal uncertainty of the HGPM, but the systematic error may have a different allocation. The required magnitude of these hypothetical small scale perturbations is approximately 70 \uasyr\ in GPM or 1.7 mas in Hipparcos positions. The only alternative option that the formal errors of Gaia proper motions are underestimated by a factor 2--3 seems even less credible.

This unexpected error budget overhead will complicate the search for astrometric long-period exoplanets from Hipparcos and Gaia proper motions. It also makes the prospect of improving the Hipparcos positional and proper motion systems using Gaia DR3 data dimmer because of the uncertainty about its origin. 

The main result of this study is the expansion of the GPM $-$ HPM differential proper motion field in 126 low-degree VSH functions revealing numerically large terms and several statistically highly significant patterns. To ensure consistency with the previous expansion, the same sample of carefully selected ``less perturbed" Hipparcos stars was used, as well as the same technical algorithm. We find vastly greater perturbations with a median vector length of 191 \uasyr. The VSH spectrum is dominated by the three magnetic terms representing rigid rotation, which together amount to a global spin of $226\pm 7$ \uasyr, or a global angular acceleration of $9.1$ $\mu$as yr$^{-2}$. The culprit in this case is quite clearly the system of Hipparcos proper motions. We recall that a system of condition equations from a Hipparcos-like astrometric mission based on differential angular measurements separated by a basic angle has an intrinsic rank-deficiency of 6, which corresponds to a 3D rotation of the positional system and a 3D spin of the proper motion system. In the Hipparcos iterative adjustment, the abscissa zero-points of the reference great circles are not constrained by the available data, so that these specific components can be arbitrarily large. An elaborate scheme of alignment to the ICRF frame was used to estimate the position rotation \citep{1997A&A...323..620K}, while the proper motion spin remained practically out of the reach. The problem with Hipparcos proper motions goes deeper, because even larger sky-correlated perturbations have been demonstrated on a smaller spatial scale beyond the coverage of this VSH analysis \citep[Makarov, 2022 in press;][]{2022AJ....164...36Z}. These imperfections will limit the absolute accuracy of the optical CRF, as a multi-epoch, quasi-inertial reference system, to the level of $\sim 1$ mas.

\bibliography{main}
\bibliographystyle{aasjournal}

\end{document}